\documentclass[12pt]{iopart}
\pdfoutput=1
\usepackage{graphicx}
\usepackage{dcolumn}
\usepackage{bm}
\usepackage{url}
\usepackage{amssymb}
 \expandafter\let\csname equation*\endcsname\relax 
  \expandafter\let\csname endequation*\endcsname\relax 
\usepackage{amsmath}
\usepackage[colorlinks=true, a4paper=true, pdfstartview=FitV,
linkcolor=blue, citecolor=blue, urlcolor=blue]{hyperref}

\include{epsf} 

\newcommand{\N}{{\rm I\! N}}
\newcommand{\R}{{\rm I\! R}}

\begin{document}

\title{Wind speed modeled as an indexed semi-Markov process}

\author{Guglielmo D'Amico}  
\address{Dipartimento di Scienze del Farmaco, 
Universit\`a "G. D'Annunzio" di Chieti-Pescara,  66013 Chieti, Italy}
\author{Filippo Petroni}
\address{Dipartimento di Scienze Economiche ed Aziendali,
Universit\`a degli studi di Cagliari, 09123 Cagliari, Italy}
\author{Flavio Prattico}
\address{Dipartimento di Ingegneria Industriale e dell'Informazione e di Economia, Universit\`a degli studi dell'Aquila, 67100 L'Aquila, Italy}
\bigskip

\date{\today}

\begin{abstract} 
The increasing interest in renewable energy, particularly in wind, has given rise to the necessity of accurate models for the generation of good synthetic wind speed data. Markov chains are often used with this purpose but better models are needed to reproduce the statistical properties of wind speed data. 
In a previous paper we showed that semi-Markov processes are more appropriate for this purpose but to reach an accurate reproduction of real data features high order model should be used. In this work we introduce an indexed semi-Markov process that is able to fit real data. 
We downloaded a database, freely available from the web, in which are included wind speed data taken from L.S.I. -Lastem station (Italy) and sampled every 10 minutes. We then generate synthetic time series for wind speed by means of Monte Carlo simulations. The time lagged autocorrelation is then used to compare statistical properties of the proposed model with those of real data and also with a synthetic time series generated though a simple semi-Markov process.
\end{abstract}

\maketitle

\section{Introduction}

Wind represents one of the most popular renewable energy. Through wind turbines it is possible to transform kinetic energy of the wind into electrical energy that we use, daily, in our applications. The rotor of the wind turbines captures the kinetic energy of the mass of air that transient into its area, this allow the rotation of the rotor and, consequentially, the production of the electrical energy. Wind turbine is a very complex system composed of many mechanical parts, electric parts and electronics control. Most of these components need, especially the mechanical ones, to be designed at fatigue failure, caused to the cyclically trend of the wind. A powerful stochastic model, that allows the generation of synthetic wind speed data, can help the design task. Sometimes, in fact, it is just the lack of unknown loads applied to the system to make difficult its design. The capability, known the value of the actual wind speed, to estimate the successive value can assist the system control, in the common wind turbine, to recover more energy by rotating the blades according to the increasing wind speed and, at the same time, maintaining constant the rotor speed.

There are many models in the literature for the generation of synthetic wind speed data or forecasting of it. Typically the approach followed are based on neural network models \cite{14,03}, autoregressive models \cite{05,07}, Markov chain \cite{sha05,nfa04,06} or also hybrid models where are used some of the previous, or also combined with common statistic distributions, but each with a specific purpose \cite{you03,01,08,10,11}. The main problem of the neural network approach is due to the necessity of manipulate original data by decomposition or normalization. This consideration can be extended also to other hybrid models that follow, partially, a neural network approach. Most of these models, including those which use simple autoregressive approach, have the aim to forecast the wind speed known previous values.
Instead, to generate synthetic data, Markov chains of first or higher order are often used \cite{sha05,nfa04,you03,01}. In particular in \cite{sha05} is presented a comparison between a first-order Markov chain and a second-order Markov chain. A similar work, but only for the first-order Markov chain, is conduced by \cite{nfa04}, presenting the probability transition matrix and comparing the energy spectral density and autocorrelation of real and synthetic wind speed data. A tentative to modeling and to join speed and direction of wind is presented in \cite{you03}, by using two models, first-order Markov chain with different number of states, and Weibull distribution. With the same purpose Catino et al. \cite{01} define channels that group wind speed of specific directions, the transition between one channel to another is managed by Markov chain while the dynamic of the wind speed into a specific channel is described by an autoregressive model. Other applications of Markov chain can be found in \cite{06,13}. The first one uses a Markov chain model with a memory index to forecast wind gust instead, in the second one is proposed a Hidden Markov model to generate synthetic wind speed data using, as observed data to find the stochastic parameters, the atmospheric pressure, the variation of which is the direct consequence of wind. A comparison with many of these model is conduced by \cite{16} in which is proposed a method, for synthetic data generation, based on wavelet transform; the synthetic wind speed time series generated is compared with those coming from two autoregressive model (first and second order), Gauss and Weibull distribution and Markov chain model.

Then, many of these models use Markov chains to generate synthetic wind speed time series but the search for a better model is still open. Approaching this issue, 
in a previous work \cite{dami12} we proposed different second order semi-Markov models for generation of synthetic time series of wind speed. We showed that, through statistical tests, semi-Markov processes are more adequate to model the process than the simpler Markov chain. We have showed also that, comparing the autocorrelations function of real and synthetic data,  semi-Markov models can reproduce feature of real data but the autocorrelation drops to zero faster than for real data. 
In order to overcome the problem of low autocorrelation, in this paper, we propose an indexed semi-Markov model for wind speed modeling \cite{dami11b,dami12b}. More precisely we assume that wind speed is described by a discrete time homogeneous semi-Markov process in which we introduce a memory index that takes into account periods of high and low wind speed. In \cite{dami11b,dami12b} we have showed that this model are able to capture the autocorrelation function of the volatility of financial return process by using a specific index process. Here we will show that, by choosing a different index process, the model is able to describe also wind speed. 

The paper is organized as follows. First of all we present the model and the equations governing the process. Then, we introduce the database used and, by applying Monte Carlo simulations, features of real and synthetic data from the models are compared.

\section{Wind speed semi-Markov model with memory}

In this section, we give a new statistical models of wind speed, and consequently we propose a novel forecasting method based on a particular indexed semi-Markov model in discrete time. Indexed semi-Markov processes have been proposed recently by \cite{dami11a} and \cite{dami11b,dami12b}. 
In the first paper different theoretical aspects are addressed while the other papers consider different indexed process which are shown to be able to describe the persistence in financial data.\\
\indent Here we consider a specific model able to capture the persistence in the wind speed data. This is an important task and model based on Markov chains have been showed to have a poor fitting of the observed autocorrelation function. To improve the fitting, recently we proposed \cite{dami12} an higher order semi-Markov chain model. With a second order semi-Markov chain the significant gain in the autocorrelation is showed still to be too far from the observed time series. Obviously, an increase of the order would improve the results but the increase in the parameters and the computational effort will be dramatically affected. For this reasons, here we consider a more parsimonious approach which reveals to perform much better in reproducing the statistical features of wind speed, especially, the autocorrelation function.\\ 

\indent Let $(\Omega,\mathbf{F},P)$ be a probability space and consider the stochastic process
\begin{equation}
J_{-(m+1)},J_{-m},J_{-(m-1)},...,J_{-1},J_{0},J_{1},...
\end{equation}
with a finite state space $E=\{1,2,...,S\}$. In our framework the random variable $J_{n}$ describes the wind speed at the $n$-th transition.\\
\indent Let us consider the stochastic process
\begin{equation}
T_{-(m+1)},T_{-m},T_{-(m-1)},...,T_{-1},T_{0},T_{1},...
\end{equation}
with values in $\N$. The random variable $T_{n}$ describes the time in which the $n$-th transition of the wind speed occurs. We denote the stochastic process $\{X_{n}\}_{n\in \N}$ where $X_{n}$  is the sojourn time in state $J_{n-1}$ before the $n$th jump. Thus we have for all $n\in \N$ $X_{n}=T_{n+1}-T_{n}$.\\
\indent Let us consider also the stochastic process
\begin{equation}
U_{-(m+1)},U_{-m},U_{-(m-1)},...,U_{-1},U_{0},U_{1},...
\end{equation}
with values in $\R$. The random variable $U_{n}$ describes the value of the index process at the $n$th transition.\\
\begin{equation}
\label{funcrela}
U_{n}^{m}=\frac{1}{T_{n}-T_{n-(m+1)}}\sum_{k=0}^{m}\sum_{s=1}^{X_{n-1-k}}f(J_{n-1-k},s),
\end{equation}
where $f:E\times \N \rightarrow \R$ is a Borel measurable bounded function and $U_{-(m+1)}^{m},...,U_{0}^{m}$ are known and non-random.\\
\indent The process $U_{n}^{m}$ can be interpreted as a moving average of the accumulated reward process with the function $f$ as a measure of the permanence reward per unit time.\\
\indent The function $f$ depends on the state of the system $J_{n-1-k}$ and on the time $s$.\\
\indent It should be noted that the order of the moving average is on the number of transitions. As a consequence, the moving average is executed on time windows of variable length.\\
\indent The indexed model is fully specified once the dependence structure between the variables is assumed. Toward this end we adopt the following assumption:
\begin{equation}
\label{kernel}
\begin{aligned}
& P[J_{n+1}=j,\: T_{n+1}-T_{n}\leq t |\sigma(J_{h},T_{h},U_{h}^{m}),\, h=-m,...,0,...,n, J_{n}=i, U_{n}^{m}=v]\\
& =P[J_{n+1}=j,\: T_{n+1}-T_{n}\leq t |J_{n}=i, U_{n}^{m}=v]:=Q_{ij}^{m}(v;t),
\end{aligned}
\end{equation}
\noindent where $\sigma(J_{h},T_{h},U_{h}^{m}),\, h\leq n$ is the natural filtration of the three-variate process.\\
\indent The matrix of functions ${\bf Q}^{m}(v;t)=(Q_{ij}^{m}(v;t))_{i,j\in E}$ is called $\emph{indexed semi-Markov}$ $\emph{kernel}$.\\
\indent The joint process $(J_{n},T_{n})$, which is embedded in the indexed semi-Markov kernel, depends on the moving average process $U_{n}^{m}$, the latter acts as a stochastic index. Moreover, the index process $U_{n}^{m}$ depends on $(J_{n},T_{n})$ through the functional relationship $(\ref{funcrela})$.\\

\indent Observe that if 
\begin{equation}
\mathbb{P}[J_{n+1}=j,\: T_{n+1}-T_{n}\leq t |J_{n}=i, U_{n}^{\lambda}=v]=\mathbb{P}[J_{n+1}=j,\: T_{n+1}-T_{n}\leq t |J_{n}=i]
\end{equation}
\noindent for all values $v\in \R$ of the index process, then the indexed semi-Markov kernel degenerates in an ordinary semi-Markov kernel and the  model becomes equivalent to classical semi-Markov chain models as presented for example in \cite{jans06}, \cite{barb08}, \cite{dami12c}.   The dependence of the process $(J_{n},T_{n})$ from the new variable $U_{n}^{m}$ is introduced in order to capture the effect of the past transitions on the future ones for those processes which are strongly autocorrelated.\\

\indent To describe the behavior of our model at whatever time $t$ we need to define additional stochastic processes.\\
\indent Given the three-dimensional process $\{J_{n}, T_{n}, U_{n}^{m}\}$ and the indexed semi-Markov kernel ${\bf Q}^{m}(v;t)$, we define by
\begin{equation}
\label{stocproc}
\begin{aligned}
& N(t)=\sup\{n\in \mathbb{N}: T_{n}\leq t\};\\
& Z(t)=J_{N(t)};\\
& U^{m}(t)=\frac{1}{t-T_{(N(t)-\theta)-m}}\sum_{k=0}^{m}\sum_{s=1}^{(t\wedge T_{(N(t)-\theta)+1-k})-T_{(N(t)-\theta)-k}}f(J_{(N(t)-\theta)-k},s),
\end{aligned}
\end{equation}
where $T_{N(t)}\leq t < T_{N(t)+1}$ and $\theta =1_{\{t=T_{N(t)}\}}$.\\
\indent The stochastic processes defined in $(\ref{stocproc})$ represent the number of transitions up to time $t$, the state of the system (wind speed) at time $t$ and the value of the index process (moving average of function of wind speed) up to $t$, respectively. We refer to $Z(t)$ as an indexed semi-Markov chain.\\
\indent The process $U^{m}(t)$ extends the process $U_{n}^{m}$ to the case where time $t$ can be a transition or a waiting time. It is simple to realize that if $\forall m$, if  $t=T_{n}$ we have that $U^{m}(t)=U_{n}^{m}$.\\  
\indent In the papers by \cite{dami11a}, \cite{dami11b} explicit renewal-type equations were given to describe the probabilistic behavior of the indexed semi-Markov process. We do not report here those results applied to our model because, in the implementation of the model given in next section we follow a Monte Carlo simulation based approach.

\section{Application to real data}
To check the validity of our model we perform a comparison of the behavior of real data
and wind speeds generated through Monte Carlo simulations based on the model described above.
In this section we describe the database of real data used for the analysis, the method used to simulate
synthetic wind speed time series and, at the end, we compare results from real and simulated data.   

\subsection{Database}
\begin{figure}
\centering
\includegraphics[height=6cm]{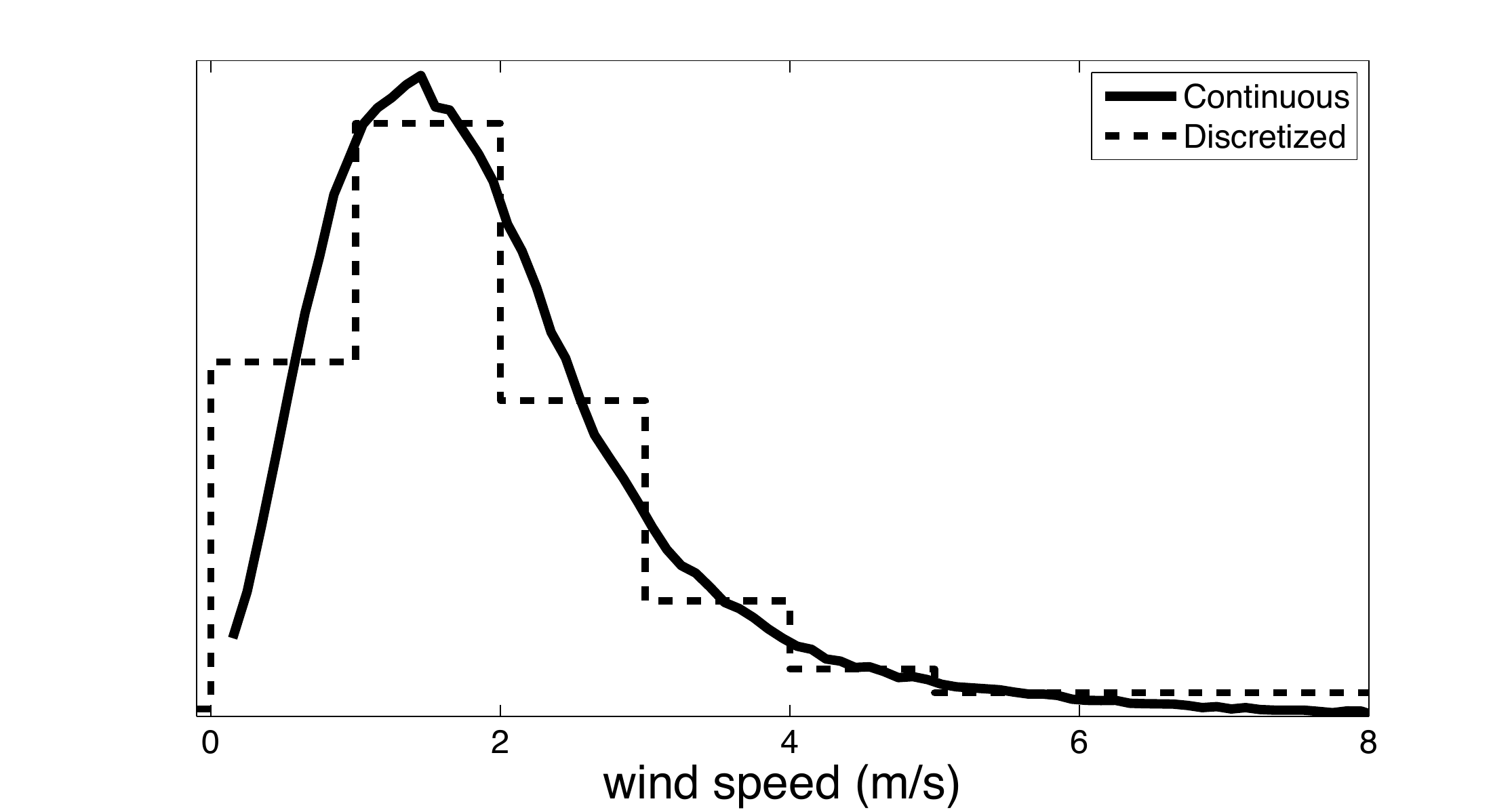}
\caption{Discretization of the wind speed distribution.}\label{discW}
\end{figure}

The data used in this analysis are freely available from $http://www.lsi-lastem.it/meteo/page/dwnldata.aspx$. The station of L.S.I. -Lastem is situated in Italy at N 45$�$ 28' 14,9'' $-$ E 9$�$ 22' 19,9'' and at 107 $m$ of altitude. The station uses a combined speed-direction anemometer at 22 $m$ above the ground. It has a measurement range that goes from 0 to 60 $m/s$, a threshold of 0,38 $m/s$ and a resolution of 0,05 $m/s$. The station processes the speed every 10 minutes in a time interval ranging from 25/10/2006 to 28/06/2011. 
During the 10 minutes are performed 31 sampling which are then averaged in the time interval.
In this work, we use the sampled data that represents the average of the modulus of the wind speed ($m/s$) without considering a specific direction.
The database is then composed of about 230thousands wind speed measures ranging from 0 to 16 $m/s$. 
To be able to model the wind speed as a semi-Markov process the state space of wind speed has been discretized.
In the example shown in this work we discretized wind speed into 8 states chosen to cover all the wind speed distribution. 
The state space is shown in Figure \ref{discW} and numerically represented by the set $E=\{0-1,1-2,2-3,3-4,4-5,5-6,6-7,>7\}(m/s)$.
This choice is done by considering a trade off between accuracy of the description of the wind speed distribution and number of parameters to be estimated.  An increase of the number of states better describes the process but requires a larger dataset to  get reliable estimates and it could also be not necessary for the accuracy needed in forecasting future wind speeds. Note also that, in the database used in this work, there are very few cases where the wind speed exceeds $7m/s$. We stress that the discretization should be chosen according to the database used. To recover continuous wind speed a simple transformation should be made \cite{dami12}. We used the following formula
\begin{equation}
w_{c}(t)=w_{d}(t)+\epsilon \Delta.
\end{equation}
where $w_{c}$ indicates the continuous wind speed at time $t$, $w_{d}(t)$ the discretized wind speed at the same time, $\epsilon$ is a uniformly distributed random number in the interval $[0,1]$ and $\Delta$ is the wind speed interval used for discretization.

\subsection{Monte Carlo simulations}

In the model described in the previous section and in particular in the definition of the index process $U^m$ the function $f:E\times \R \rightarrow \R$ is any Borel measurable bounded function. To perform simulations, we choose a function which is motivated essentially by simplicity, we want to keep the model as simple as possible. 

Let us briefly remind that wind speed data are long range positively autocorrelated. This implies that there are periods of high and low speed. Motivated by this empirical facts we suppose that also the transition probabilities depends on whether the wind is, on average, in a high speed period or in a low one.  In \cite{dami11b,dami12b} to reproduce the autocorrelation of the square of returns in financial markets we used, as index function, a moving average of the square of returns. Given that in this case is the wind speed itself to be correlated we decided to choose a moving average of the wind speed. 
We then fixed the function $f$ to be the wind speed itself, i.e. 
\begin{equation}\label{f}
f(J_{n-1-k},s)=J_{n-1-k} 
\end{equation}
for all $s\in \N$. Consequently, substituting (\ref{f}) in equation (\ref{funcrela}) and considering that $J_{n-1-k} $ is constant in $s$ we obtain 
\begin{equation}
\label{indicespecifico}
\begin{aligned}
U_{n}^{m}=& \frac{1}{T_{n}-T_{n-(m+1)}}\sum_{k=0}^{m}J_{n-1-k}\cdot X_{n-1-k} \\
& =\sum_{k=0}^{m} J_{n-1-k}\bigg(\frac{T_{n-k}-T_{n-1-k}}{T_{n}-T_{n-(m+1)}}\bigg)
\end{aligned}
\end{equation}
\indent In this simple case the index process expresses a moving average of order $m+1$ executed on the series of the wind speed values with weights given by the fractions of sojourn times in that wind speed, with respect to the interval time on which the average is executed.\\

Note that the memory is the number of transitions. The index $U^m$ obtained from the given definition of $f$ was also discretized into 5 states of low, medium low, medium, medium high and high speed.
 
According to these choices we estimated, from real data, the probabilities $Q_{ij}^m(v;t)$ defined in formula (\ref{kernel}) for different values of $m$. For the results shown below $m$ was chosen to run from 1 to 30 transitions.

Then, these probabilities have been used to simulate synthetic time series of wind speed \cite{code}  needed to compare results from real data and the model, as will be described in the next section. 
Here we give a Monte Carlo algorithm in order to simulate the trajectories in the time interval $[0, T]$. The algorithm consists in repeated random sampling to compute successive visited states of the random variables $\{J_{0}, J_{1},...\}$, the jump times $\{T_{0},T_{1},...\}$ and the index values $\{U_{0}^{m},U_{1}^{m},... \}$ up to the time $T$.\\
The algorithm consists of 5 steps:\\
1) Set $n=0$, $J_{0}=i$, $T_{0}=0$, $U_{0}^{m}=v$, horizon time$=T$;\\
2) Sample $J$ from $p_{J_{n},\cdot}^{m}(U_{n}^{m})$ and set $J_{n+1}=J(\omega)$;\\
3) Sample $W$ from $G_{J_{n},J_{n+1}}^{m}(U_{n}^{m},\cdot)$ and set $T_{n+1}=T_{n}+W(\omega)$;\\
4) Set $U_{n+1}^{m}=\frac{1}{T_{n+1}-T_{n+1-(m+1)}}\sum_{k=0}^{m}\sum_{s=1}^{X_{n-k}}f(J_{n-k},s)$\\
5) If $T_{n+1}\geq T$ stop\\
\indent else set $n=n+1$ and go to 2).

\begin{figure}
\centering
\includegraphics[height=8cm]{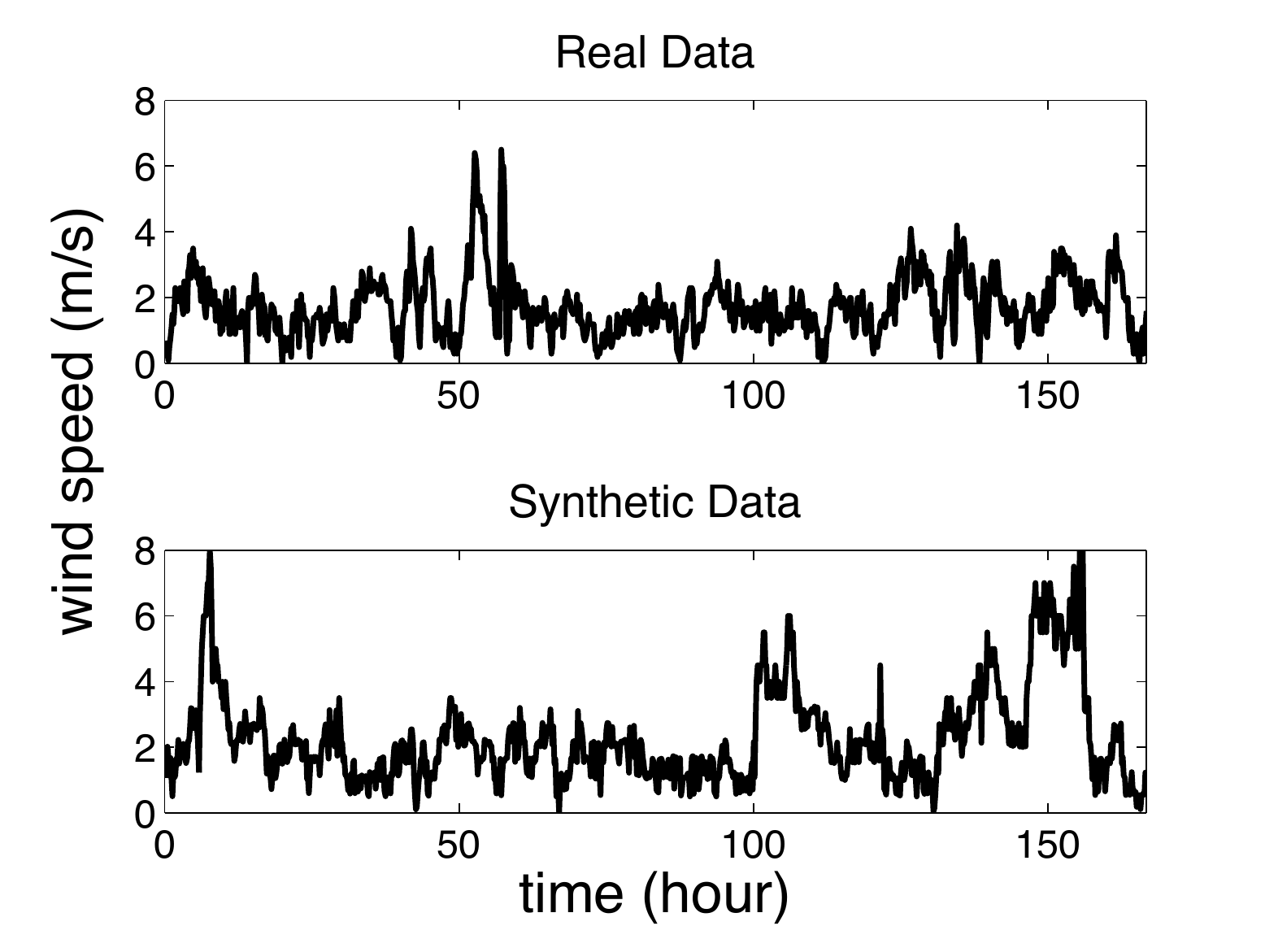}
\caption{Trajectories of real and simulated data.}\label{tra}
\end{figure}

An example of the trajectories of real and simulated data is shown in Figure \ref{tra}.
Note that these are step-by-step simulations in which the index $U^m$ has to be calculated from the last $m$ simulated transitions.

\subsection{Comparison of real and simulated data}
A very important feature of wind speed data is that  they are long range correlated. It is then very important that theoretical models do reproduce this features.  We tested our model to check whether it is able to reproduce such behavior. 
Given the presence of the parameter $m$ in the index function, we also tested the autocorrelation behavior as a function of $m$.

If $Z$ indicates wind speed, the time lagged $(\tau)$ autocorrelation of wind speed is defined as 
\begin{equation}
\label{autosquare}
\Sigma(\tau)=\frac{Cov(Z(t+\tau),Z(t))}{Var(Z(t))}
\end{equation}
We tested our model on two different time scale of wind speed, one in which data are measured every 10 minutes and one in which data are measured every 1 hour. The time lag $\tau$ was made to run from 1 minute up to 1000 minutes in the first case and from 1 hour to 50 hours in the second one. Note that to be able to compare results for $\Sigma(\tau)$ each simulated time series was generated with the same length as real data.
\begin{figure}
\centering
\includegraphics[height=5.2cm]{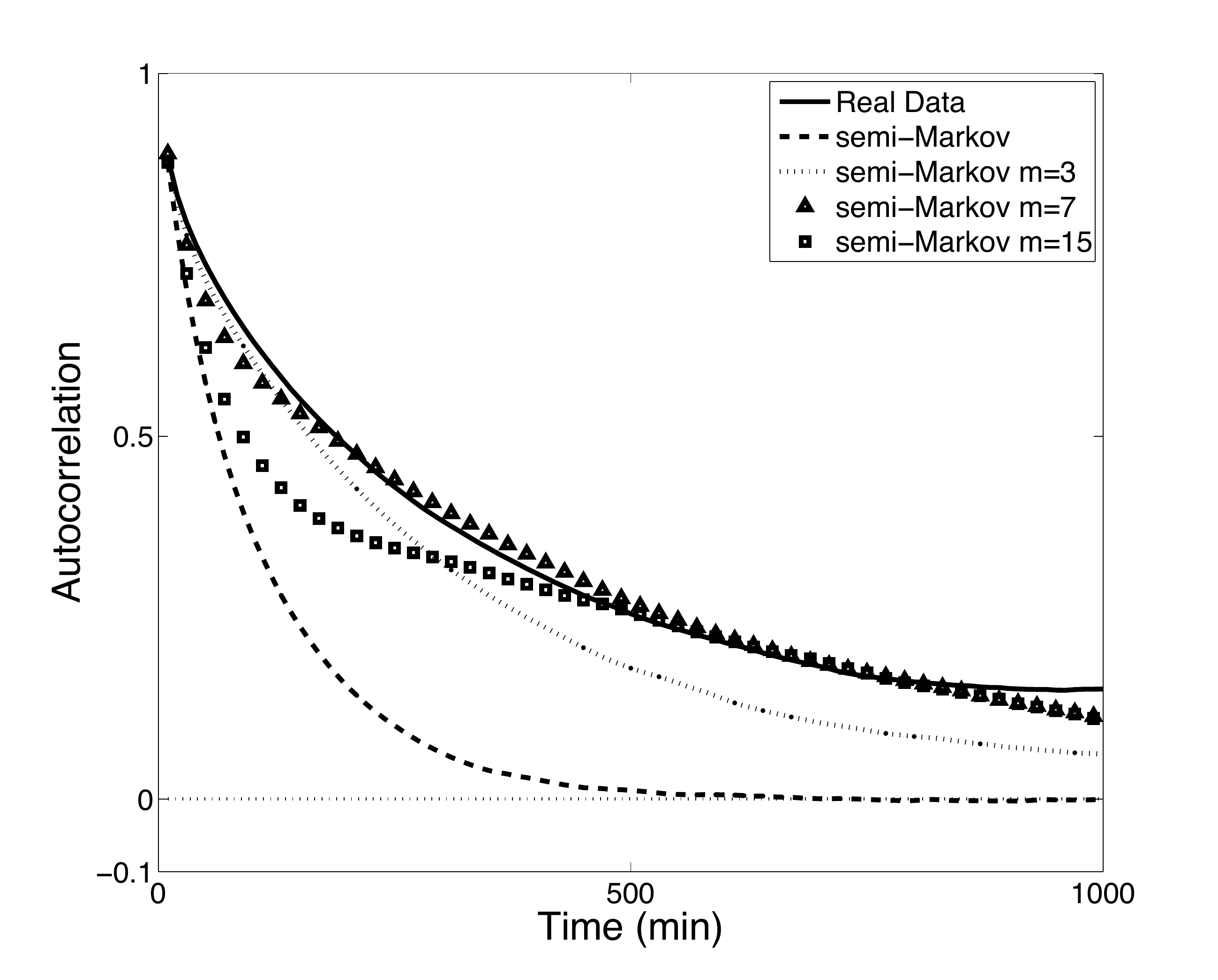}
\includegraphics[height=5.2cm]{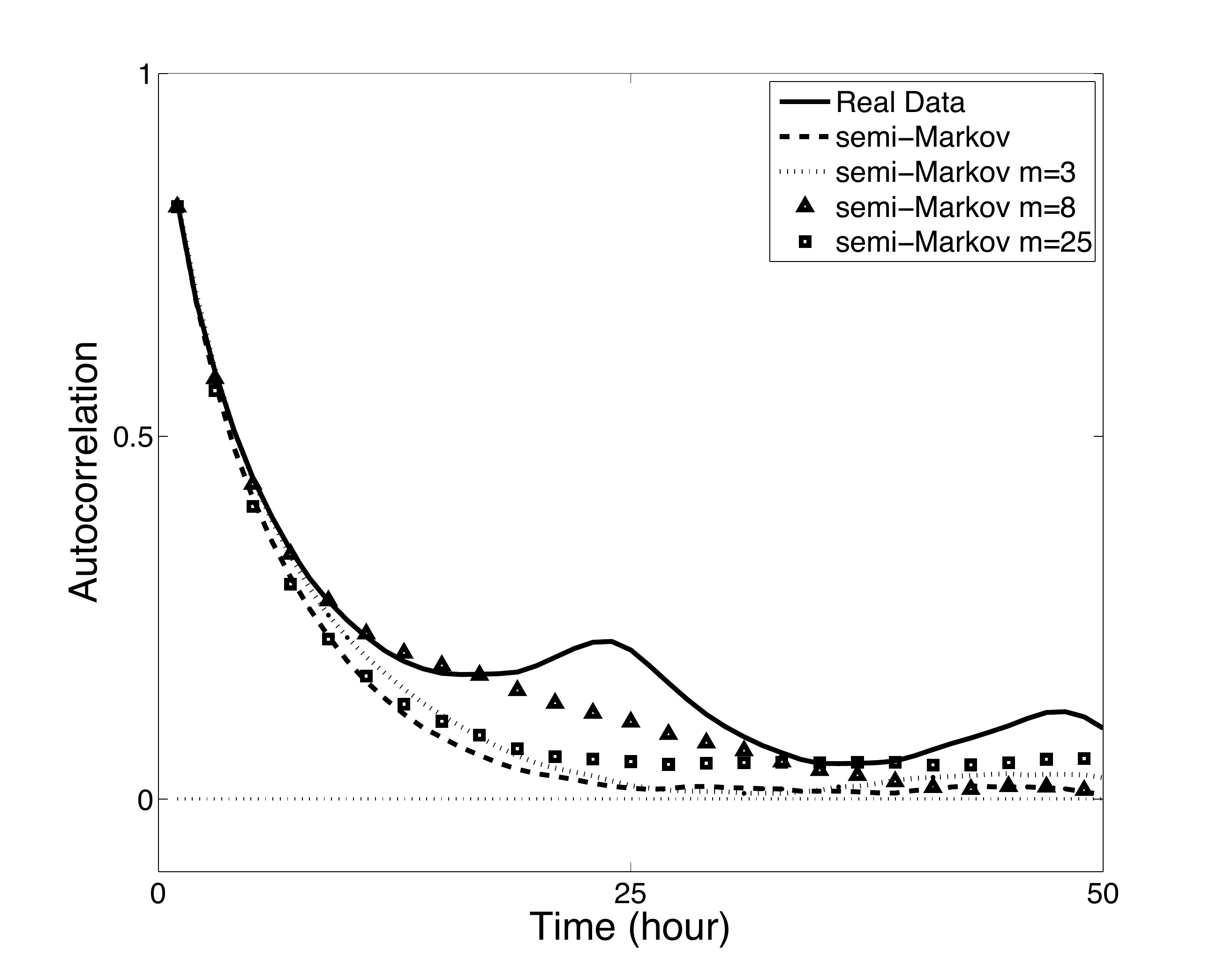}
\caption{Autocorrelation functions of real data (solid line) and of 4 synthetic time series as described in the label. In the left panel wind speed is measured every 10 minutes while in the right panel every hour.}\label{fig2}
\end{figure}
As expected (see Figure \ref{fig2}), real data do show a long range correlation with a sinusoidal behavior, the latter is due to simple seasonal effect of the 24hour earth cycle. 
Let us then analyze results for the synthetic time series. The simple semi-Markov model starts at the same value but the persistence is very short and after few time steps the autocorrelation decreases to zero. A very interesting behavior is instead shown by the semi-Markov models with memory index. If a small memory ($m=3$ in the shown example) is used, the autocorrelation is already persistent but again decreases faster than real data. With a longer memory ($m=7/8$) the autocorrelation remain high for a very  long period and also its value is very close to that of real data. If $m$ is increased further the autocorrelation drops again to small values. This behavior suggest the existence of an optimal memory $m$. In our opinion one can justify this behavior by saying that short memories are not enough to identify in which speed status is the wind, too long memories mix together different status and then much of the information is lost in the average. 
All this is shown in Figure \ref{fig3} where the mean square error between each autocorrelation function of simulated time series and the autocorrelation function of the real data as a function of $m$ is computed. It can be noticed that there exist an optimal value of the memory $m$ that makes the autocorrelation of simulated data closer to that of real data. 
\begin{figure}
\centering
\includegraphics[height=5.2cm]{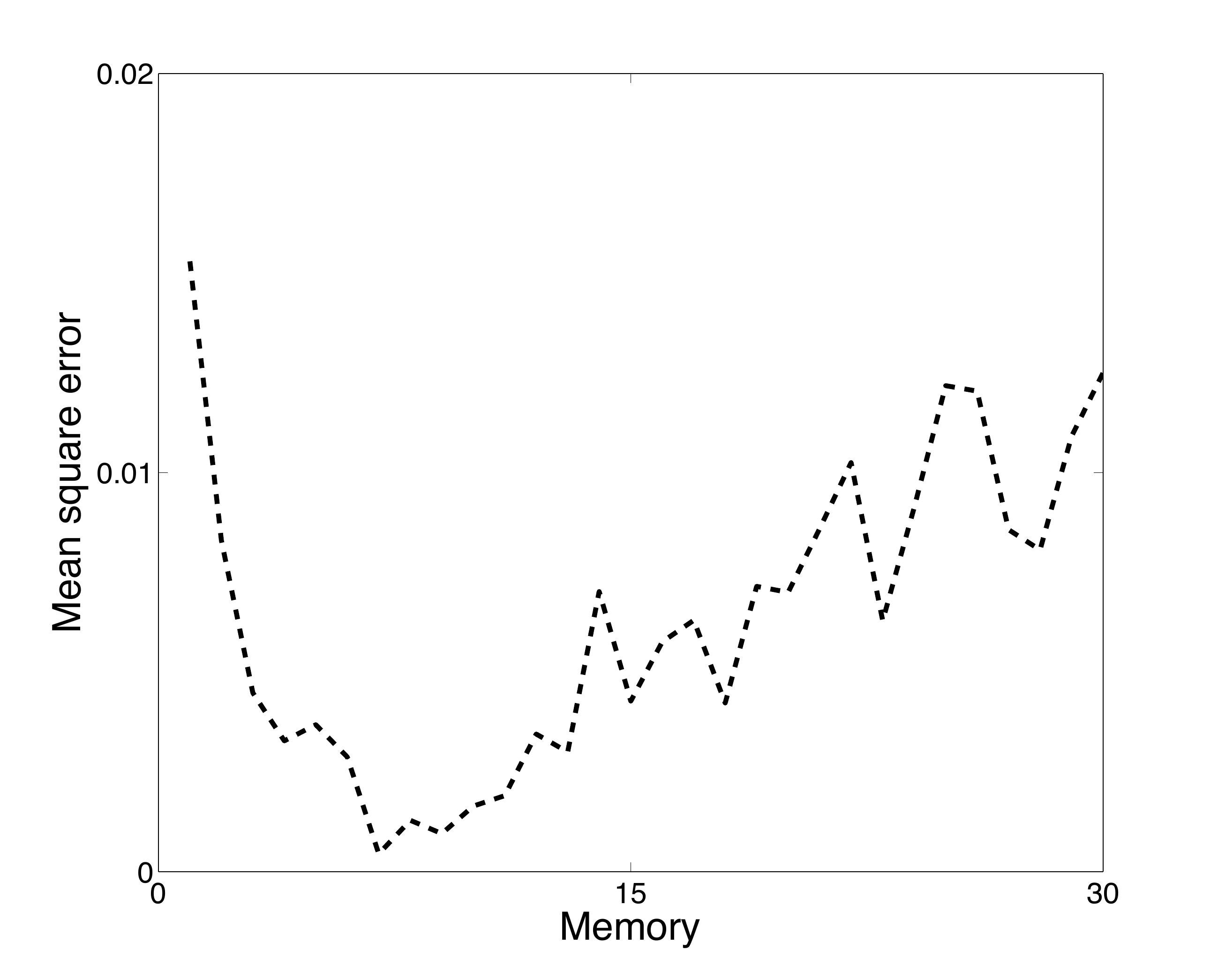}
\includegraphics[height=5.2cm]{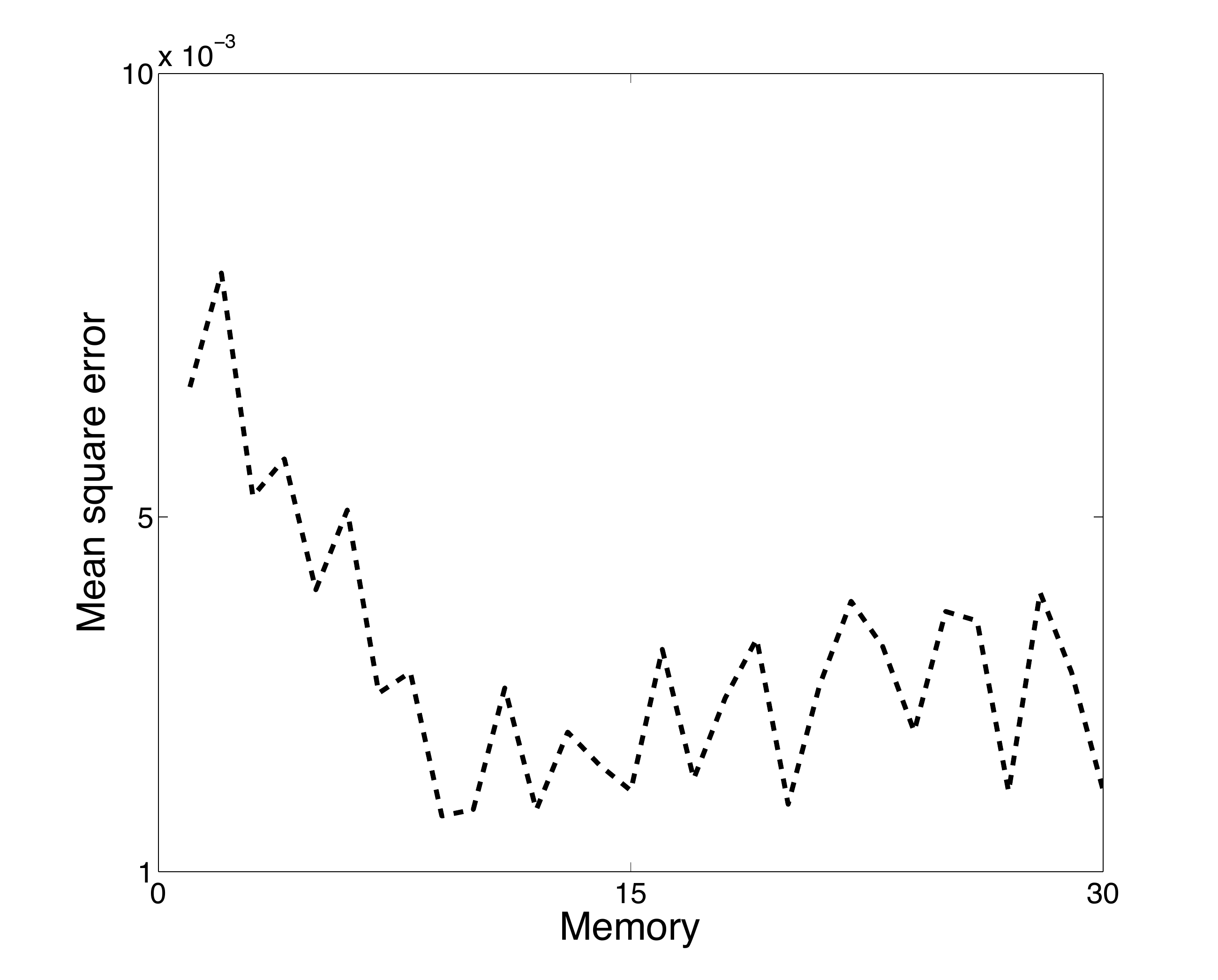}
\caption{Mean square error between autocorrelation function of real data and synthetic data as a function of the memory value $m$. As in Figure \ref{fig2} the left panel wind speed is measured every 10 minutes while in the right panel every hour.}\label{fig3}
\end{figure}
Notice that Figure \ref{fig2}, right panel, reproduce the autocorrelation function except for the seasonal term which is not included in our model.  We decided to model only the stochastic term present in the data also to show that the autocorrelation function does not depend on the periodic term which could be added easily by standard procedures.
In Figure \ref{fig4} we also compare the whole probability density function of real and simulated data showing that they are almost identical.
\begin{figure}
\centering
\includegraphics[height=8.2cm]{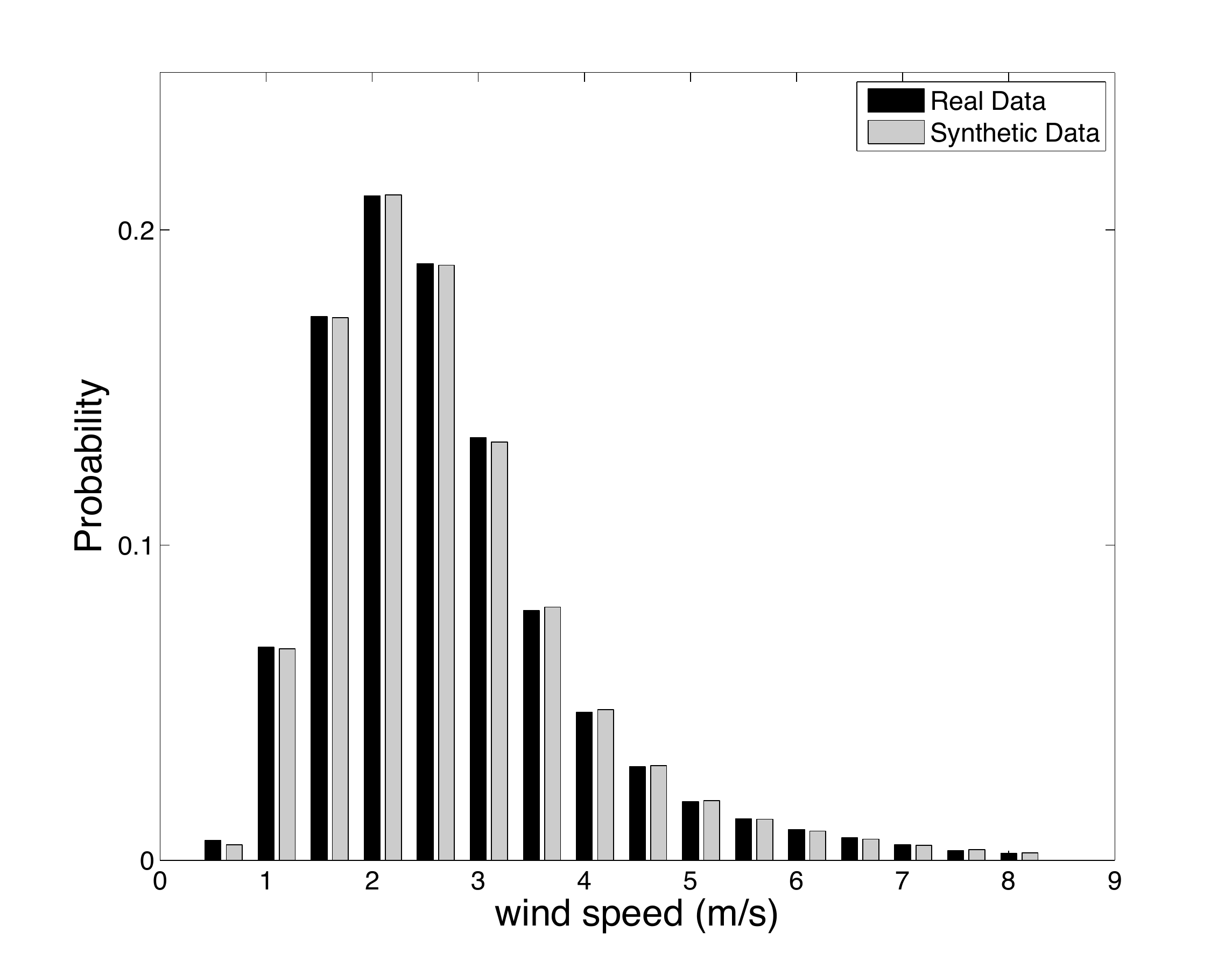}
\caption{Probability density function for real and synthetic data.}\label{fig4}
\end{figure}

\section{Conclusion}
The wind is a very unstable phenomenon characterized by a sequence of lulls and
sustained speeds, and a good stochastic model of wind speed must be able to reproduce such
sequences. We have then modeled wind speed through a semi-Markov model where we have added a memory index. Our work is motivated by the presence of persistence in the wind speed process.

The results presented here show that the semi-Markov kernel is influenced by the past wind speed. In fact, if the past wind speed is used as a memory index, the model is able to reproduce quite well the statistical properties of real data.

We have also shown that the time length of the memory does play a crucial role in reproducing the right autocorrelation's persistence indicating the existence of an optimal value. 

We modeled wind speed not considering the periodic 24hours components which can be added easily if the model is to be used for forecasting at lags greater than 24 hours.
The main features of our approach are the following: the first is that since our model is completely non  parametric it does not require assumption to be made about the data. Second, it generalizes models based on Markov chains which are very often used for solving autoregressive processes numerically.  It is able to reproduce data with high persistence in a natural way without the use of exogenous variables and also by using a limited number of states.

The work leaves many open issues like for example: the determination of more general guidelines to select the index function, practical applications of the model regarding energy production, blades and turbines design.

\end{document}